\documentclass[aps,prb,showpacs,twocolumn,amsmath,amssymb]{revtex4}

\usepackage{graphicx}
\usepackage{dcolumn}
\usepackage{bm}
\usepackage{psfrag}
\usepackage{subfigure}

\newcommand{\bra}[1]{\langle #1 |}
\newcommand{\ket}[1]{| #1 \rangle}
\newcommand{\braket}[2]{\langle #1 | #2\rangle}

\newcommand{\Eq}[1]{Eq.~(\ref{#1})}
\newcommand{\Eqs}[1]{Eqs.~(\ref{#1})}

\begin{document}


\title{Manipulation with Andreev states in spin active mesoscopic Josephson junctions
}%

\author{J. Michelsen}
\author{V.S. Shumeiko}
\author{G. Wendin}
\affiliation{%
Department of Microtechnology and Nanoscience, MC2\\
Chalmers University of
Technology, SE-41296 Gothenburg, Sweden\\
}%

\date{\today}

\begin{abstract}

We investigate manipulation with Andreev bound states in  Josephson
quantum point contacts with magnetic scattering. Rabi oscillations
in the two-level Andreev subsystems are excited by resonant driving
the direction of magnetic moment of the scatterer, and by
modulating the superconducting phase difference across the contact.
The Andreev level dynamics is manifested by temporal oscillation of
the Josephson current, accompanied, in the case of magnetic
manipulation, also by oscillation of the Andreev states spin
polarization. The interlevel transitions obey a selection rule that
forbids manipulations in a certain region of external parameters,
and results from specific properties of Andreev bound states in
magnetic contacts: 4$\pi$-periodicity with respect to the
superconducting phase, and strong spontaneous spin polarization.

\end{abstract}

\pacs{74.50.+r, 74.45.+c, 71.70.Ej}%
\maketitle


\section{Introduction}
Recent advances in the development and experimental investigation
of nanowire based Josephson junctions
\cite{Delft,Copenhagen,Grenoble} attract new attention to rich
physics of mesoscopic Josephson effect. One of particularly
interesting questions concerns the possibility to employ Josephson
quantum point contacts for quantum information processing. Such
contacts contain a small number of generic two-level systems -
Andreev bound levels, whose quantum states can be selectively
manipulated and measured.\cite{Shumeiko93,Gorelik95}  By modulating
the phase difference across the junction one is able  to induce the
Rabi oscillation in the Andreev two-level system, and therefore to
prepare arbitrary superposition of the Andreev states. A
measurement of induced oscillation of the Josephson current allows
for the Andreev level readout. Thus the pair of Andreev bound
levels belonging to the same conducting mode may serve as a quantum
bit.\cite{ALQ,ALQ2}

An interesting possibility to involve a spin degree of freedom in
the contact quantum dynamics, and to use it for qubit application
has been investigated by Chtchelkatchev and
Nazarov.\cite{Nazarov03} They considered a Josephson quantum point
contact with spin-orbit interaction and showed how to manipulate
with the spin of the Andreev state. The properties of Andreev bound
states in spin-active mesoscopic junctions and equilibrium
Josephson effect have been extensively studied in recent
literature;\cite{Fogelstrom00,Fogelstrom01,Barash02a,Barash02b,Bezuglyi02,Radovic03,Vecino03,Radovic06,Benjamin07}
non-stationary aspects of interaction of individual magnetic
scatterers with Josephson current have also been
discussed.\cite{Zhu03,Zhu04,Bulaevskii04,Nussinov05}

In this paper, we investigate the methods of manipulation with the
Andreev states in Josephson quantum point contacts containing a
magnetic scatterer, e.g. magnetic nanoparticle situating between
the superconducting
electrodes.\cite{Grenoble,Wernsdorfer97,Kasumov05} We investigate
two manipulation methods: (i) time variation of the superconducting
phase across the contact, and (ii) time variation of the  direction
of magnetic moment of the  scatterer. We find that in both the
cases the Josephson current exhibits  Rabi oscillation under the
resonant drive within a certain interval of biasing superconducting
phase. In the case of magnetic manipulation, the effect may only
exist if the Andreev states are initially spin polarized, then the
current oscillation is accompanied by oscillation of the Andreev
states spin polarization. The phase interval, where the Rabi
oscillation can be excited, decreases with increasing strength of
the magnetic scatterer, and eventually disappears at large enough
strength; in particular in the $\pi$-junction regime, the Rabi
oscillation is completely forbidden. This selection rule results
from specific properties of the bound Andreev states in magnetic
junctions as we will show.

\begin{center}
\begin{figure}[h!]
\includegraphics[width=0.4\textwidth]{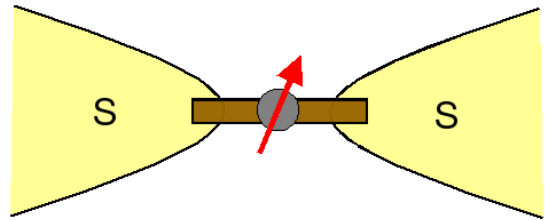}
\caption{Sketch of a magnetic Josephson point contact:
superconducting reservoirs are connected by a nanowire of the
length $L$ smaller than the coherence length, magnetic nanoparticle
creates local classical magnetic field. }\label{Contact}
\end{figure}
\end{center}

For a static scatter the spin rotation symmetry around the
direction of its magnetic moment  is preserved. This allows for
 the contact description in terms
of a two-component Nambu spinor,\cite{Atland97} similar to
non-magnetic junctions, thus avoiding a double counting problem.
Within such an approach, the two bound Andreev levels per
conducting modes are only relevant, giving complete quantitative
description of the stationary Josephson effect as well as the
resonant two-level transitions and non-stationary current response
under the phase manipulation.\cite{ALQ}  Consideration of the spin
conjugated Nambu spinor gives a completely equivalent physical
description in terms of a reciprocal pair of Andreev bound states;
both the pictures mirror each other.

Time variation of the direction of the magnetic moment of the
scatterer leads to a violation of the spin rotation symmetry, and
induces coupling between the spin conjugated Nambu spinors. This
results in a unitary rotation in the extended space of the four
Andreev bound states. It turns out, however, that this rotation
splits into two equivalent rotations in invariant two-level
subspaces, which mirror each other. Thus the contact response can
also in this case be  explained in terms of the two-level Rabi
dynamics. For the two different ways of magnetic manipulation
considered - instant switching, and small-amplitude resonant
oscillation of the direction of the magnetic moment of the
scatterer, the Andreev two-level  dynamics has a physical meaning
of precession, and nutation of the spin polarization of the Andreev
levels, respectively. Thus the spin polarization of Andreev states
is required in order to observe a non-trivial dynamical
response.\cite{Nazarov03} Such a possibility naturally exists, as
we will show, in the contact under consideration: the equilibrium
Andreev states exhibit strong spin polarization, up to the maximum
values $\pm1/2$ at low temperature, in certain regions of the
 superconducting and Zeeman phases.

\section{Contact description }

Consider one-mode quantum point contact with superconducting
electrodes connected by a normally conducting nanowire as shown in
Fig. \ref{Contact}. The left and right electrodes (L,R)  are
described with the BCS Hamiltonian,
\begin{equation}\label{HS}
\begin{split}
H_S=\int_{\mathrm{L,R}} dx \sum_{\sigma=\uparrow,\downarrow}
\hat{\psi}^\dag_{\sigma}(x)  \left({p^2\over 2m} - \mu(x,t)\right)
\hat{\psi}_{\sigma}(x) \\%
+ \Delta^*(x,t)\hat{\psi}_{\uparrow}(x) \hat{\psi}_{\downarrow} (x)
+\Delta(x,t)\hat{\psi}^\dag_{\downarrow}(x)\hat{\psi}^\dag_{\uparrow}(x).
\end{split}
\end{equation}
Aiming to investigate the effect of time variation of the
superconducting phase difference $\varphi$ across the junction, we
consider the order parameter having the form, $\Delta(x,t) = \Delta
e^{i\varphi(x,t)/2}$, $\varphi(x,t) = \varphi(t)\,\mathrm{sgn}
\,x$, and the electrochemical potential, $\mu(x,t)$, having the
form, $\mu(x,t) = E_F-\hbar\,\partial_t\varphi/2$, which provides
the electro-neutrality condition within the electrodes.\cite{ALQ2}

To explore the spin properties of the Andreev states, we assume
that the contact nanowire contains a magnetic scatterer, e.g.,
magnetic nanoparticle (Fig. \ref{Contact}). We assume for
simplicity that the magnetic field ${\bf H}(x,t)$ induced by the
scatterer is localized within the nanowire on a distance $l$
smaller than the distance $L$ between the electrodes, thus not
affecting the superconductivity within the electrodes; furthermore,
we will treat this magnetic field as a given external parameter
neglecting the back action effect from the current. Then the
Hamiltonian of the normal region of the junction has the form,
\begin{equation}\label{HN}
\begin{split}
H_N = \int_{-L/2}^{L/2} dx \sum_{\sigma\sigma'}
\hat{\psi}^\dag_{\sigma}(x) \left[\left({p^2\over 2m} - E_F +
U(x)\right)\delta_{\sigma\sigma'}\right.\\
\left. + {1\over 2}\mu_B {\mbox{\boldmath$\sigma$}}_{\sigma\sigma'}
\mathbf{H}(x,t)\right] \hat{\psi}_{\sigma'}(x),
\end{split}
\end{equation}
where $U(x)$ is the scalar potential of the scatterer. We assume a
symmetric with respect to $x=0$ spatial distributions of the scalar
potential and the magnetic field, and also a fixed direction of the
magnetic field, which will vary with time during the manipulation.

In the stationary case, the Hamiltonian (\ref{HN}) preserves the
spin rotational symmetry around direction of the magnetic field.
Choosing the spin quantization axis along this direction, we
describe the electron propagation through the normal region of the
junction with a transfer matrix, $T_e$,
\begin{equation}\label{S}
T_e=\hat{d}^{-1}%
\begin{pmatrix}%
 e^{i\sigma_z(\beta/2)} & i\hat{r}\\
-i\hat{r} & e^{-i\sigma_z(\beta/2)}
\end{pmatrix},
\end{equation}
where $\hat d = \rm{diag} (d_{\uparrow},  d_{\downarrow})$ and
$\hat r = \rm{diag}(r_{\uparrow},  r_{\downarrow})$.  Contact
description in terms of the spin active scattering matrix has been
extensively discussed in literature.\cite{Barash02a,Lofwander04}
The impurity scalar potential produces spatially symmetric
scattering with transmission amplitudes,
$d_{\uparrow},d_{\downarrow}$, and reflection amplitudes,
$r_{\uparrow},r_{\downarrow}$, which may be different for different
spin orientations (spin selection). The scattering phase shift
$\beta$ between the opposite spin orientations is induced by the
Zeeman effect,
\begin{equation}\label{beta}
\beta = {\mu_B Hl\over \hbar v_F}.
\end{equation}

The physical observables of interest, Josephson current and spin
polarization, are described with a single electron density matrix
$\rho(x,x',t)$ associated with a two-component Nambu
field\cite{Nambu60} $\Psi(x,t)$,
\begin{equation}\label{rho}
\Psi(x,t) =
\begin{pmatrix} \hat\psi_\uparrow(x,t) \\ \hat\psi_\downarrow^\dagger(x,t)
 \end{pmatrix}, \quad %
 \rho(x,x',t) = \langle \Psi(x,t)\Psi^\dagger(x',t)\rangle
\end{equation}
(here  the angle brackets indicate statistical averaging).
Alternative description is given by a density matrix,
$\tilde\rho(x,x',t)$ associated with the spin conjugated Nambu
field, $\tilde\Psi(x,t)$,
\begin{equation}\label{tildePsi}
\tilde\Psi(x,t) =
\begin{pmatrix} \hat\psi_\downarrow(x,t) \\ -\hat\psi_\uparrow^\dagger(x,t)
 \end{pmatrix}, \quad \tilde\rho(x,x',t)=\langle
\tilde\Psi(x,t)\tilde\Psi^\dagger(x',t)\rangle.
\end{equation}
The two Nambu fields are connected via a fundamental symmetry
relation imposed by the singlet nature of the BCS pairing,
\begin{equation}\label{symmetryPsi}
\tilde\Psi(x,t)= i\sigma_y\left(\Psi^\dag\right)^T.
\end{equation}

In what follows we will use these two reciprocal representations,
$\phi$-representation and $\tilde\phi$-representation, respectively
(see Appendix B). This will allow us to avoid the redundancy,
introduced by \Eq{symmetryPsi}, of a commonly used four-component
Nambu
formalism,\cite{Fogelstrom00,Fogelstrom01,Barash02a,Barash02b,Bezuglyi02,Radovic03,Radovic06}
and to explicitly show that the stationary Josephson effect as well
as non-stationary response to the phase manipulation can be fully
understood within the framework of the two bound Andreev states per
conducting mode associated with either of these two reciprocal
representations.

\section{Andreev state subsystem}

In this section we outline the properties of the stationary Andreev states,
which are of importance for the discussion of non-stationary effects. The
details of derivations are presented in Appendix A.

Within the two-component Nambu formalism,\cite{Nambu60} the
quasiparticle states in the stationary contact are given by the
Bogoliubov-deGennes equation,
\begin{equation}\label{BdG}
h\phi = E\phi, \quad h= (p^2/2m - E_F)\,\sigma_z +
\Delta\,\sigma_x,
\end{equation}
supplemented with the boundary condition at the contact, which for a short
contact is given by the transfer matrix, $T=\exp(i\sigma_z\varphi/2)T_e$,
\Eq{T_BC}. The energy spectrum of the bound states consists of the two
levels, according to \Eq{ALspectrumApp},
\begin{equation}\label{ALspectrum}
E_{s}=\,\theta_s\Delta\cos(s\eta-\beta/2),\;\quad s=\pm,
\end{equation}
where $\theta_s=\mathrm{sgn}[\sin(s\eta-\beta/2)]$,  and parameter
  $\eta$ is defined through equation, $\sin\eta = \sqrt
D\sin(\varphi/2)$, $D$ is the contact transparency. This energy
spectrum is asymmetric with respect to the chemical potential,
$E=0$, and is $4\pi$-periodic with respect to the phase difference
as shown on Fig. \ref{spectrum}, the spectral branches cross at
$\varphi= 2\pi n$. On the figure are also shown the energy levels
of a reciprocal Andreev level pair, $\tilde E_s = - E_s$, whose
wave functions, $\tilde\phi_{s}=(i\sigma_y)\phi_{s}^\ast$, are
associated with the spin conjugated Nambu field, \Eq{tildePsi}
($\tilde\phi$-representation,  see Appendix B).

\begin{center}
\vspace{0.1cm}
\begin{figure}[h]
\includegraphics[width=0.45\textwidth]{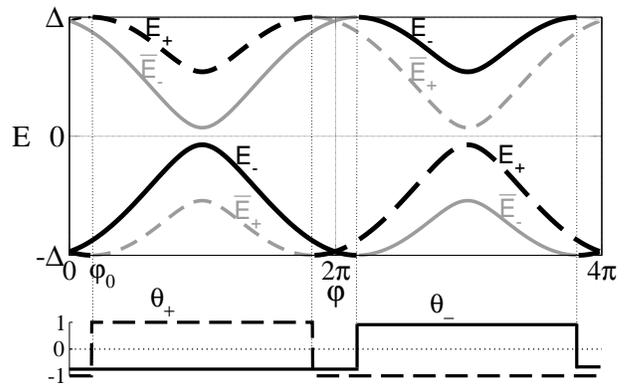}
\caption{Andreev level energy spectrum for $\sin(\beta/2)<\sqrt D$
($D$=0.9, $\beta$=0.5),  upper panel; $\theta$-functions, shown on
the lower panel, define the spectrum discontinuity points at $2\pi
n\pm \varphi_0$. Vertical dotted lines separate regions of weak and
strong Zeeman effect.  }\label{spectrum}
\end{figure}
\end{center}

The reason for the $4\pi$-periodicity and the level crossings is
the symmetry of the problem which is reflected in the property,
\Eq{symmetry}. Introducing the parity operator $P$ that permutes
the wave functions at the left and right sides of the junction,
$P\phi(x<0)= \phi(x>0)$, we write \Eq{symmetry} on the
form,\cite{Gorelik98}
\begin{equation}\label{symmetry2}
\hat\Lambda\hat{\phi}_s(x) = s\theta_s\hat{\phi}_s(x), \quad \hat\Lambda = P
\begin{pmatrix}%
- \sigma_x & 0\\
0 & \sigma_x
\end{pmatrix}.
\end{equation}
According to this equation, the Andreev level wave functions are
simultaneously the eigen functions of the symmetry operator $\hat\Lambda$
with eigenvalues $s\theta_s$. Furthermore, the energy branches, $E_-$ and
$E_+$, in the neighboring phase intervals in Fig. \ref{spectrum}, correspond
to different eigen values of the operator $\Lambda$. The properties of the
Andreev states are therefore qualitatively different within the  phase
intervals, where $\theta_s = \theta_{-s}$, and $\theta_s = -\theta_{-s}$. To
emphasize the difference we will refer to the former ones  as the regions of
strong Zeeman effect (ZE), $\sin\beta/2> \sqrt{D}\sin\varphi/2$, and the
latter ones as the regions of weak ZE, $\sin\beta/2< \sqrt{D}\sin\varphi/2$.
These regions are separated by the points where the energy levels touch the
continuum, $\varphi = 2\pi n \pm\varphi_0$,
$\varphi_0=2\arcsin(1/\sqrt{D}\sin\beta/2)$. At sufficiently large Zeeman
phase, $\sin\beta/2>\sqrt{D}$ the weak ZE regions disappear, while strong ZE
regions spread over the whole superconducting phase axis. The bound energy
levels, $E_\pm(\varphi)$, depart from the continuum forming "cigars", see
Fig.~\ref{cigars} , they belong to the orthogonal eigen subspaces of the
operator $\hat\Lambda$ at all phases. The $\pi$-contact is realized in this
regime, at $\beta=\pi$,\cite{Kulik65} when the reciprocal cigars coincide and
situate symmetrically with respect to $E=0$.

\begin{center}
\vspace{0.1cm}
\begin{figure}[h]
\includegraphics[width=0.4\textwidth]{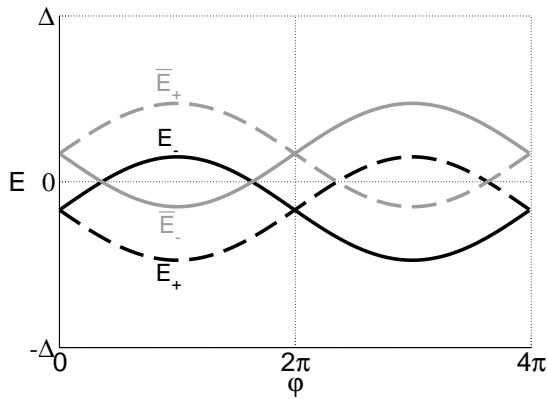}
\caption{Andreev level energy spectrum for $\sin(\beta/2)>\sqrt D$
($D$=0.1, $\beta=2.8$). Strong ZE region spreads over whole phase
axis.}\label{cigars}
\end{figure}
\end{center}

A qualitative difference between the regions of strong and weak ZE is
illustrated by the properties of Josephson current. Starting with a general
expression for the charge current through the density matrix,
\begin{equation}\label{qcurrent}
I(t)=\frac{e\hbar}{2mi}\left(\partial_x-\partial_{x'}
\right)\left[\delta(x-x') - \textrm{Tr}\rho(x,x',t)\,
\right]_{x=x'=0},
\end{equation}
we truncate it to the Andreev level subspace,
\begin{equation}\label{Irho}
I_A(t)=\sum_{ss'}I_{ss'}\left(\frac{1}{
2}\,\delta_{s's}-\rho_{s's}(t)\right).
\end{equation}
Here
\begin{equation}\label{rhoss'}
\rho_{ss'}(t)=\bra{\phi_s}\rho(t)\ket{\phi_{s'}}=\int dx dx'
\textrm{Tr}\left[\rho(x,x',t)\phi_{s'}(x')\phi_s^\dag(x)\right],
\end{equation}
is the density matrix in the Andreev level representation,
\Eqs{psiR}, (\ref{symmetry}), the trace refers to the electron-hole
space. The current matrix $I_{ss'}$ reads,
\begin{equation}\label{Iss'weak}
\begin{split}
 I_{ss'}=&\frac{2e}{\hbar} \left[
\begin{pmatrix}
\partial_\varphi E_+ & 0\\
0 & \partial_\varphi E_-
\end{pmatrix} \right.\\
 +\,&\left. (1-\theta_s\theta_{-s})\sqrt{RD}\sin\frac{\varphi}{2}
\frac{\sqrt{\zeta_+\zeta_-}} {2\varepsilon}
\begin{pmatrix} 0 & 1\\ 1 & 0
\end{pmatrix}\right].
\end{split}
\end{equation}
The diagonal elements give the expectation values of the currents
of Andreev states, while the off-diagonal part describes the
current quantum fluctuation.\cite{ALQ2} The off-diagonal part is
finite in the weak ZE regions, where $\theta_s\theta_{-s}=-1$,
while it is zero in the strong ZE regions. This implies that the
current quantum fluctuation is fully suppressed in the strong ZE
regions: here the current matrix commutes with the Andreev level
Hamiltonian, which is diagonal in this representation $h_{ss'} =
\delta_{ss'}E_s$.

In the equilibrium, $\rho_{ss'}=\delta_{ss'}n_F(-E_s)$, and the
Andreev  current in \Eq{Irho} takes the  form,
\begin{equation}\label{eq.andreevcurrent}
\begin{split}
I_A = \,-\,{e\Delta\over\hbar} \sum_{s}\partial_\varphi E_s \,
 \tanh(E_s/2T),
\end{split}
\end{equation}
One can confirm by direct calculation that the current of the
continuum states vanishes, thus  \Eq{eq.andreevcurrent} represents
the total Josephson current. This equation coincides with the
result obtained using the four-component Nambu
formalism,\cite{Fogelstrom00,Fogelstrom01,Barash02a,Barash02b,Bezuglyi02,Radovic03,Radovic06}
and it can be also obtained by working with the reciprocal Nambu
representation  using equation (\ref{tilde}).

Spin polarization of the Andreev states plays an important role in
magnetic manipulation. Asymmetry of the Andreev spectrum with
respect to the zero energy together with the spectrum
discontinuities result in a peculiar phase dependence of the spin
polarization of  Andreev levels. The $z$-component of electronic
spin density in the contact is given by equation,
\begin{equation}\label{sdensity}
S(x,t)={1\over 2}\left[\delta(x-x') - \textrm{Tr}\rho(x,x',t)\,
\right]_{x=x'}.
\end{equation}
Truncating this equation to the Andreev level subspace, and
integrating over $x$ using normalization condition for the bound
state wave functions, we find the spin polarization of the Andreev
level pair,
\begin{equation}\label{andreevspin}
S_A = (1/ 2)(1 - f_+ - f_- ),
\end{equation}
where $f_s=\rho_{ss}$ are the population numbers.

\begin{center}
\begin{figure}[h]
\includegraphics[width=0.4\textwidth]{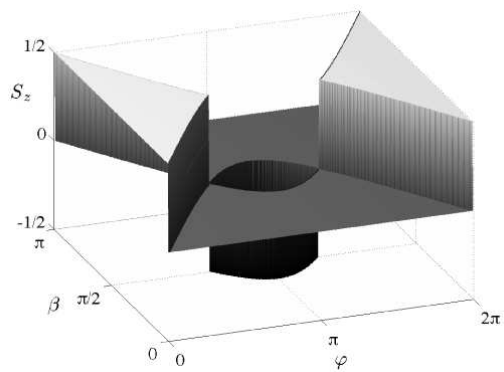}
\caption{Equilibrium spin polarization of Andreev levels at zero
temperature and at $D>1/2$; the central region with $S=-1/2$
emerges at $D=1/2$ at $\varphi=\pi,\,\beta=\pi/2$, and grows with
$D$ reaching the lines $\beta = 0$ and $\beta=\pi$ at $D=1$.
}\label{Spin}
\end{figure}
\end{center}

Thus the spin polarization of the Andreev levels is entirely determined by
their (generally non-equilibrium) population numbers: For empty Andreev level
pair, $f_\pm=0$, the spin polarization is $S= 1/2$, while for fully populated
levels, $f_\pm=1$, it is $S=-1/2$. For the single particle occupation of the
level pair, $f_+ +f_- =1$, the spin polarization is zero, $S=0$. In
non-magnetic contacts the equilibrium spin polarization of  Andreev levels is
always zero, $S=0$, by virtue of the identity, $n_F(-E_+)+n_F(-E_-)=1$, that
holds due to the spectrum symmetry, $E_+= - E_-$. In magnetic contacts, the
spin polarization sharply varies in the $\varphi-\beta$ parameter plane, see
Fig.~\ref{Spin}. At zero temperature, $f_s = \theta(E_s)$, the spin
polarization is zero in the regions bound by the lines, $\sin(\varphi/2) =
\sin(\beta/2)/\sqrt D$ and $\sin(\varphi/2) = \cos(\beta/2)/\sqrt D$; the
first region corresponds to a weak ZE. Outside these regions the Andreev
levels are strongly polarized, $S=1/2$. At small contact transparencies,
$D<1/2$, the above mentioned lines do not overlap, while at $D>1/2$ they do,
forming the island of strong negative polarization, $S=-1/2$, around the
point, $\varphi=\pi$, $\beta=\pi/2$, see Fig.~\ref{Spin}. This island grows
with increasing transparency eventually touching the lines, $\beta=0$ and
$\beta=\pi$ at $D=1$.

In contrast to the charge density, the spin density in
\Eq{sdensity} obeys the conservation equation,  $\partial_t
S(x,t)+\partial_x I_S(x,t)=0$. The consequence of this is the zero
spin current of the Andreev states, $I_{SA}=0$: under the
stationary condition the partial spin current of the Andreev state
must be constant in space, and being proportional to the bound wave
function it vanishes at the infinity; thus it is identically equal
to zero.\cite{comment}

\section{Phase manipulation}

\begin{center}
\begin{figure}[h]
\includegraphics[width=0.4\textwidth]{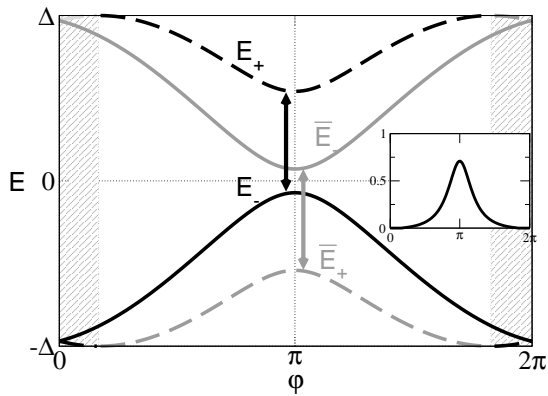}
\caption{Interlevel transitions induced by time oscillation of the
phase difference, shadow regions indicate forbidden regions;
transitions in $\tilde\phi$ representation (reduced-intensity
lines) are equivalent to the transitions in $\phi$-representation
(full-intensity lines); $D$=0.9, $\beta$=0.5. Inset: transition
matrix element as function of $\varphi$. }\label{Phasemanipulation}
\end{figure}
\end{center}

Now we turn to discussion of the Andreev level dynamics under the
time dependent phase. Similar to  non-magnetic junctions this
dynamics involve only the bound levels belonging to the same Nambu
representation,\cite{Shumeiko93,Gorelik95,ALQ} as shown on Fig.
\ref{Phasemanipulation}. The frequency of the phase time variation
must be small compared to the distance to the gap edges to prevent
the level-continuum transitions, $\omega \ll \Delta - |E_s|$.

Time evolution of the contact density matrix, $\rho(t)$, is governed by the
Liouville equation, $i\hbar
\partial_t\rho =[h,\rho]$, with the
Hamiltonian of \Eq{BdG}, supplemented with the non-stationary boundary
condition, $T=\exp(i\sigma_z\varphi(t)/2)T_e$. Now we truncate  the density
matrix using the instantaneous Andreev eigenfunctions, $\phi_s[\varphi(t)]$
(cf. \Eq{rhoss'}),
\begin{equation}
\rho_{ss'}(t)=\bra{\phi_s(t)}\rho(t)\ket{\phi_{s'}(t)}.
\end{equation}
This density matrix obeys the Liouville equation $i\hbar
\partial_t\rho =[H,\rho]$ with a truncated Hamiltonian, which in this basis
is given by equation,
\begin{equation}
H_{ss'}=\bra{\phi_s} h \ket{\phi_{s'}} -\bra{\phi_s} i\hbar
\partial_t \ket{\phi_{s'}}=E_s\delta_{ss'} -\hbar \dot{\varphi}
\bra{\phi_s} i
\partial_{\varphi}\ket{\phi_{s'}},
\end{equation}
where $E_s[\varphi(t)]$ and $\phi_s[\varphi(t)]$ are given by Eqs.
(\ref{ALspectrumApp}) and (\ref{psiR}), (\ref{symmetry}) respectively.

The matrix element,
$\bra{\phi_s}i\partial_{\varphi}\ket{\phi_{s'}}$, is found to be
zero for $s'=s$, while for the interlevel transitions, $s'= - s$,
it reads,
\begin{equation}\label{phiMatrixElement}
\begin{split}
\bra{\phi_s} i\partial_{\varphi} \ket{\phi_{-s}} = is
(1-\theta_s\theta_{-s}){\Lambda(\varphi)\over 2},\\
\Lambda(\varphi)= \frac{\sqrt{R}D}{2}\sin^2\frac{\varphi}{2}
\frac{\sqrt{\zeta_+\zeta_-}}{\varepsilon^2(\zeta_++\zeta_-)}.
\end{split}
\end{equation}
From this we find that in the weak ZE regions,
$\theta_s\theta_{-s}=-1$, the non-stationary Andreev level
Hamiltonian has the form,
\begin{equation}\label{ALhamiltonian}
H(t)=\begin{pmatrix} E_+ & 0\\ 0 & E_-
\end{pmatrix} + \, \hbar\dot{\varphi}\Lambda(\varphi)
\begin{pmatrix} 0 & -i\\ i & 0
\end{pmatrix}, \; \sqrt{D}\sin{\varphi\over2}>\sin{\beta\over2}.
\end{equation}
This equation provides generalization to a magnetic junction of the
Hamiltonian derived in Ref.~[\onlinecite{ALQ}]: the level coupling coincides
with the one in non-magnetic junction when $\beta=0$, and remains finite when
$\beta\neq0$ but only inside the weak ZE region, decreasing towards the edges
of this region, see inset in Fig. \ref{Phasemanipulation}. In the strong ZE
region the matrix element is identically zero ($\theta_s\theta_{-s}=+1$), and
the Hamiltonian is diagonal,

\begin{equation}
H(t)=\begin{pmatrix} E_+ & 0\\ 0 & E_-
\end{pmatrix}, \quad \sqrt{D}\sin{\varphi\over2}<\sin{\beta\over2}.
\end{equation}
Thus we conclude that no operation with Andreev levels is possible
in the strong ZE regime.

\Eq{ALhamiltonian} is convenient for calculation of Rabi
oscillation in the Andreev level system under the resonant driving,
$\varphi(t)=\varphi +\delta\sin\omega t$,  $\omega = (E_+ -
E_-)/\hbar$. Inserting this in \Eq{ALhamiltonian}, we get
\begin{equation}\label{}
\begin{split}
H &= {\hbar\omega\over 2}\sigma_z + \hbar\Lambda\delta\omega
\cos\omega t\,
\sigma_y. \\
\end{split}
\end{equation}
Assuming small amplitude of the phase oscillation, $\delta\ll 1$,
and using rotating wave approximation, we find the time dependent
density matrix in the rotating frame,
\begin{equation}\label{Rabirho}
\begin{split}
\rho(t) = \rho(0) - {f_+(0)-f_-(0)\over 2}
\,(1-\cos\Lambda\delta\omega t -
\sin\Lambda\delta\omega t\, \sigma_x), \\
\end{split}
\end{equation}
where $\rho(0) = diag (f_+(0), f_-(0))$.  Rabi oscillation of the
Andreev levels generate a time dependent Josephson current,
\begin{equation}\label{RabiI}
I(t)=I(0)-\frac{2e}{\hbar}\left(f_+(0) -
f_-(0)\right)\sin^2{\Lambda\delta\omega t\over2}\,
 \sum_s \partial_\varphi E_s\,.
\end{equation}

The Rabi oscillation, \Eq{Rabirho}, and the time oscillation of the
Josephson current, \Eq{RabiI}, vanish if the Andreev levels are
initially fully spin polarized, $S(0)=\pm1/2$, since $f_+(0) =
f_-(0)$ in this case. Thus an additional requirement for the phase
manipulation is to bias the contact in the region outside the
negative-spin island on Fig.~\ref{Spin}.

\section{Spin manipulation}

\begin{center}
\begin{figure}[h!]
\includegraphics[width=0.4\textwidth]{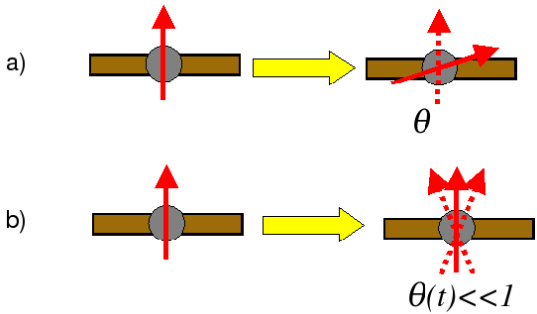}
\caption{Sketch of manipulation with magnetic field; a) instant
switching of direction of magnetic field; b) small-amplitude
oscillation of magnetic field with resonant frequency (electron
spin resonance)}\label{Hmanipulation}
\end{figure}
\end{center}

Now we proceed with the discussion of the spin manipulation. We
consider the two ways of driving Andreev level spin, presented in
Fig. \ref{Hmanipulation}: (i) rapid change of the {\em direction}
of the magnetic moment of the scatterer (dc pulsing), and (ii)
harmonic oscillation with resonance frequency of the magnetic
moment direction (rf-pulsing). In both cases the spin rotation
symmetry is violated, therefore the junction dynamics cannot be
described with only one Nambu pseudo-spinor, but involves both the
spin conjugated Nambu pseudo-spinors. The interlevel transitions in
this case physically describe a rotation of the Andreev level spin.

%
\subsection{dc pulsing}

Suppose the Andreev levels are initially prepared in a stationary state with
non-zero spin, which points along the applied magnetic field ($z$-axis). Such
states were discussed in previous sections. Let us now suppose that the
magnetic field is rapidly rotated by angle $\theta$ around $y$-axis, as shown
on Fig.~\ref{Hmanipulation}a. Such a manipulation is described by rotation of
the electronic $T$-matrix, in \Eq{S},
\begin{equation}\label{U}
T_e \rightarrow U T_e U^\dag, \quad %
U=\begin{pmatrix} \cos\frac{\theta}{2} & \sin\frac{\theta}{2} \\
-\sin\frac{\theta}{2} & \cos\frac{\theta}{2}
\end{pmatrix},
\end{equation}
and it mixes the Nambu pseudo-spinors $\Psi$ and $\tilde\Psi$. To
describe the effect of this manipulation, we introduce the extended
four-component Nambu space, $(\Psi,\tilde\Psi)^T$, and the
corresponding single particle density matrix,
\begin{equation}\label{Pi}
\Pi(x,x',t) =
\begin{pmatrix}
\langle\Psi(x,t)\Psi^\dag(x',t)\rangle &
\langle\Psi(x,t)\tilde\Psi^\dag(x',t)\rangle\\
\langle\tilde\Psi(x,t)\Psi^\dag(x',t)\rangle &
\langle\tilde\Psi(x,t)\tilde\Psi^\dag(x',t)\rangle
\end{pmatrix}.
\end{equation}
This density matrix operates in the Hilbert space spanned by the
extended eigen basis,
\begin{equation}\label{eq.wavefunctionsfour}
\Phi_\nu(x)=\begin{pmatrix} \phi_\nu(x)\\ 0
\end{pmatrix},  \quad
\Phi_{\tilde{\nu}}(x)=\begin{pmatrix} 0\\ \tilde{\phi}_\nu(x).
\end{pmatrix}.
\end{equation}
The transformation $U$ induces rotation of the extended basis
\begin{equation}
\Phi_\alpha(x) \rightarrow U\Phi_\alpha(x), \qquad
\alpha\in\{\nu,\tilde{\nu}\}.
\end{equation}
The eigen energies, however, remain the same since such a rotation
just corresponds to changing the spin quantization axis.

The wave functions, \Eq{eq.wavefunctionsfour}, form a complete set
in the extended space, and thus any operator can be expressed
through them,
\begin{equation} A(x,x') = \sum_{\alpha,\beta}\Phi_\alpha(x)
\Phi^\dag_{\beta}(x')A_{\alpha\beta}.
\end{equation}
In particular, for a stationary system with spin rotational
invariance we have for the density matrix,
\begin{equation}\label{Pialpha_beta}
\Pi(x,x')=\sum_{\alpha}\Phi_\alpha(x)\Phi^\dag_{\alpha}(x')f_{\alpha},
\end{equation}
and the Hamiltonian,
\begin{equation}
H(x)=\sum_{\alpha}\Phi_\alpha(x)\Phi^\dag_{\alpha}(x)E_\alpha .
\end{equation}
However, one has to remember that this description is redundant,
and rigorous constraints hold on the occupation numbers,
$\tilde{f}_{\nu}=1-f_\nu$,
and eigen energies,
$\tilde{E}_{\nu}=-E_\nu$.

We  now write down the Hamiltonian after the magnetic field
rotation on the form,
\begin{equation}\label{rotatedH}
H(x)=\sum_{\alpha}U\Phi_\alpha(x) E_\alpha
\Phi^{\dag}_\alpha(x)U^\dag, \qquad t>0.
\end{equation}
In the initial basis, this Hamiltonian is represented with the
matrix,
\begin{equation}
H_{\alpha\beta} = \sum_{\mu} \bra{\Phi_{\alpha}} U\ket{\Phi_\mu}
E_{\mu}\bra{\Phi_\mu} U^\dag\ket{\Phi_\beta}, \qquad t>0,
\end{equation}
or  explicitly,
\begin{equation}\begin{split}
H_{\nu\nu'}&= \cos^2\frac{\theta}{2}
E_\nu\delta_{\nu\nu'}- \sin^2\frac{\theta}{2}
\sum_{\mu\neq\nu,\nu'} \braket{\phi_\nu}{\tilde{\phi}_\mu}
E_{\mu}\braket{\tilde{\phi}_\mu}{\phi_{\nu'}}\\
H_{\tilde{\nu}\tilde{\nu}'} &=-\cos^2\frac{\theta}{2} E_{\nu}\delta_{\nu\nu'}
+\sin^2\frac{\theta}{2} \sum_{\mu\neq\nu,\nu'}
\braket{\tilde{\phi}_\nu}{\phi_\mu}
E_\mu\braket{\phi_\mu}{\tilde{\phi}_{\nu'}}\\
 H_{\nu\tilde{\nu}'}
&=- \sin\theta\braket{\phi_\nu}{\tilde{\phi}_{\nu'}}
\frac{E_\nu+E_{\nu'}}{2}.
\end{split}
\end{equation}
Here the orthogonality relations were used,
$\braket{\phi_\nu}{\phi_\mu}
=\braket{\tilde{\phi}_\nu}{\tilde{\phi}_{\mu}} =\delta_{\mu\nu}$,
and $\braket{\phi_\nu}{\tilde{\phi_\nu}}=0$,
\Eq{tildeorthogonality}.

\begin{center}
\begin{figure}[h]
\includegraphics[width=0.4\textwidth]{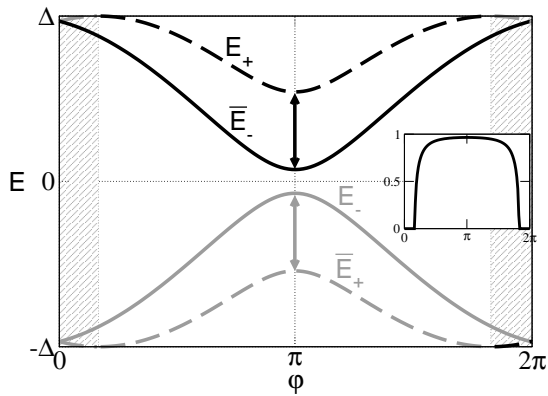}
\caption{Interlevel transitions induced by magnetic manipulation;
shadow regions indicate forbidden regions; transitions between the
levels ($E_-,\tilde E_+$)(reduced-intensity lines) are equivalent
to the transitions ($E_+,\tilde E_-$) (full-intensity lines);
$D$=0.9, $\beta$=0.5. Inset: transition matrix element as function
of $\varphi$. }\label{SpinManip}
\end{figure}
\end{center}

At this point we restrict ourselves to the Andreev level subspace,
and present the truncated Hamiltonian on the form (using the
symmetries $E_s=-\tilde{E}_s$ and
$\braket{\phi_+}{\tilde{\phi}_-}=-\braket{\phi_-}{\tilde{\phi}_+}$),
\begin{equation}\label{eq.form4hamiltonian}
H^{(4)}=\begin{pmatrix} E_0 + W &
0 & 0 & V\\
0 & -E_0 + W & -V & 0\\
0 & -V & -E_0 -W & 0 \\
V & 0 & 0 & E_0 - W
\end{pmatrix},
\end{equation}
where
\begin{equation}
\begin{split}
E_0,W &=\left(\cos^2\frac{\theta}{2} -
\sin^2\frac{\theta}{2}|M|^2\right)\,\frac{E_+
\mp E_{-}}{2}\,,\\
V &=- M\sin\theta \,\frac{E_++E_-}{2}\,,
\end{split}
\end{equation}
and the interlevel matrix element,
\begin{equation}\label{M}
M=\braket{\phi_+}{\tilde{\phi}_-} =
\left(\theta_s-\theta_{-s}\right) \cos\frac{\beta}{2}
\frac{\sqrt{\zeta_+\zeta_-}}{\zeta_++\zeta_-}\,.
\end{equation}
The matrix element (\ref{M}) equals to zero in the strong ZE region
($\theta_s =\theta_{-s}$), thus the manipulation does not produce
any effect there, which is similar to the phase manipulation, see
inset in Fig. \ref{SpinManip}.

The Hamiltonian (\ref{eq.form4hamiltonian}) has a block-diagonal
form describing identical rotations in the two orthogonal
subspaces, spanned with the eigen vectors
($\phi_+,\,\tilde\phi_-$), and ($\phi_-,\,\tilde\phi_+$). Thus the
problem reduces to solving for two physically equivalent two-level
systems. Choosing the subspace ($\phi_+,\,\tilde\phi_-$), we have
the two-level Hamiltonian,
\begin{equation} H^{(2)}=\begin{pmatrix} W & V\\ V & -W
\end{pmatrix}.
\end{equation}
Introducing projection operators on the eigen subspaces,
\begin{equation}
\begin{split}
H^{(2)} &= \sum_{\lambda=\pm} \lambda \hbar\Omega P_\lambda, \quad
P_\lambda = {1\over2}\left(1 + \lambda {\sigma_z W +\sigma_x V
\over \hbar\Omega}\right),
\\
\hbar\Omega &= \sqrt{W^2+V^2}=\frac{(E_++E_-)}{2}
\left(\cos^2\frac{\theta}{2} +\sin^2\frac{\theta}{2}|M|^2\right),
\end{split}
\end{equation}
where $\lambda\hbar\Omega$ are the eigen energies, we have for the
time evolution of the two-level density matrix,
\begin{equation}\label{}
\Pi^{(2)}(t) = e^{-i\frac{Ht}{\hbar}}\Pi^{(2)}(0)
e^{i\frac{Ht}{\hbar}}=\sum_{\lambda\lambda'}
e^{i(\lambda'-\lambda)\Omega t} P_\lambda \Pi^{(2)}(0)
P_{\lambda'}.
\end{equation}
Assuming the initial density matrix to be stationary (not
necessarily equilibrium), and expressing it through the level
occupation numbers of the $\phi$-representation,
\begin{equation}\label{eq.density2}
\Pi^{(2)}(0)=
\begin{pmatrix}
f_+(0) & 0\\
0 & 1-f_-(0)
\end{pmatrix},
\end{equation}
we obtain,
\begin{equation}\label{Pi2}
\begin{split}
 \Pi^{(2)}(t)= \Pi^{(2)}(0) +2S_A(0)
\begin{pmatrix}
|a(t)|^2 & b(t)\\
 b(t)^\ast & - |a(t)|^2
\end{pmatrix},\\
a(t) = -i\frac{V}{\hbar\Omega}\sin\Omega t, \quad
 b(t)=a(t)\left[\cos\Omega t+ i\frac{W}{\hbar\Omega}\sin\Omega t\right],
\end{split}
\end{equation}
where $S_A(0)=(1/2)[1-f_+(0)-f_-(0)]$ is the initial
spin-polarization of the Andreev levels as given by
\Eq{andreevspin}. Thus no rotation is induced for spin unpolarized
Andreev levels. Furthermore, the frequency of the rotation is
proportional to the level splitting, $\hbar\Omega \propto E_+ -
\tilde E_- = E_+ + E_-$, i.e. to the magnetic field.

The time-evolution of the occupation numbers $f_s(t)$ of the
Andreev levels in the $\phi$-representation is extracted from
\Eq{Pi2},
\begin{equation}\label{f_s}
f_s(t)=f_s(0) + 2S_A(0)\frac{V^2}{(\hbar\Omega)^2}\sin^2\Omega t.
\end{equation}
This relation illustrates the non-unitary evolution of the Andreev
levels in this representation, $f_+(t)+f_-(t)\neq$ const.

\Eq{f_s} allows us to  obtain the time dependence of the spin
polarization of the Andreev levels,
\begin{equation}\label{eq.magneticmomenttime}
S_A(t)={1\over2}(1-f_+(t)-f_-(t)) = S_A(0) \left(1 -
\frac{2V^2}{(\hbar\Omega)^2}\sin^2\Omega t\right).
\end{equation}

To calculate the Josephson current, we use the expression through
the current matrix in the $\phi$-representation, \Eqs{Irho},
(\ref{Iss'weak}),
\begin{equation}\label{}
I(t)=\sum_{ss'}I_{ss'}\left(\frac{1}{
2}\delta_{s's}-\rho_{s's}(t)\right).
\end{equation}
 The diagonal elements of the density matrix,
$\rho_{ss}$, are given by \Eq{f_s}. On the other hand, the
off-diagonal elements, $\rho_{ss'}$, equal zero because the spin
manipulation does not induce transitions between the eigen states
of the same (either $\phi$- or $\tilde\phi$-) representation.
Therefore,
\begin{equation}
\begin{split}
I(t) &={2e\over\hbar}\sum_s\partial_\varphi E_sf_s(t) \\
&= I(0)+{4e\over\hbar}
S_A(0)\frac{V^2}{(\hbar\Omega)^2}\sin^2\Omega
t\sum_{s}\partial_\varphi E_s.
\end{split}
\end{equation}

Summarizing, the conditions for the observation of the
non-stationary contact response, biasing in a weak ZE region with a
finite spin polarization, can be only fulfilled in the central
island region on Fig.~\ref{Spin} with negative polarization. This
constraint can be relaxed by pumping initial level populations away
from equilibrium as suggested in Ref.\cite{Nazarov03}.

 Concluding this section we note that the dc pulsing of the
magnetic field does not allow one to reach every point on the Bloch
sphere: this is due to the fact that $U$ is not an invariant
operation on the Andreev level subspace of the extended Nambu
space, which physically means a leakage to the continuum (similar
effect exists also for the phase manipulation with dc
pulses\cite{Lantz02}). However, for small rotation angles and not
close to the edges of the weak ZE region,  the matrix element $M$
is close to unity, and the leakage is small; it can be further
reduced by using rapid adiabatic change of the magnetic field, i.e.
rapid on the time scale of the Andreev level splitting but slow on
the time scale of the distance of the Andreev levels to the
continuum. This shortcoming does not exist for the resonant
rf-pulsing.
\subsection{rf-pulsing}

Now let us consider a time-dependent rotation of the magnetic
field,
\[T\rightarrow U(\theta(t))TU^\dag(\theta(t)).\]
where $\theta(t)$, as before, is the angle of rotation around the $y$-axis,
see Fig.~\ref{Hmanipulation}b. We can now define the instantaneous
eigenstates
\begin{equation}\label{eq.timedepU}
U(t)\Phi_\alpha(x),
\end{equation}
satisfying the instantaneous boundary condition, \Eq{S},  with
$T_e\rightarrow UT_eU^\dag$. The time-dependent Hamiltonian can
then be written similarly to \Eq{rotatedH},
\begin{equation}
H(x,t)=\sum_{\alpha}U(t)\Phi_{\alpha}(x)E_{\alpha}\Phi_\alpha^\dag(x)U^\dag(t).
\end{equation}
Since the energy eigenvalues do not depend on the direction of the
quantization axis they remain time-independent. Now similarly to
\Eq{Pialpha_beta} we can expand the density matrix in terms of
these instantaneous eigenfunctions,
\begin{equation}
\Pi(x,x',t)=\sum_{\alpha,\beta}
U(t)\Phi_\alpha(x)\Phi^\dag_\beta(x')U^\dag(t)\Pi_{\alpha\beta}(t).
\end{equation}
The matrix $\Pi_{\alpha\beta}(t)$ satisfies the Liouville equation
with the Hamiltonian,
\begin{equation}\begin{split}
H_{\alpha\beta}(t)=&\int dx
\Phi^\dag_{\alpha}(x)U^\dag(t)\left[H(x)-i\hbar\partial_t\right]U(t)\Phi_\beta(x)\\
  =& E_{\alpha}\delta_{\alpha\beta} - i\hbar\bra{\Phi_\alpha}
U^\dag(t) \partial_tU(t)\ket{\Phi_\beta}.
\end{split}
\end{equation}
Inserting Eq. (\ref{eq.timedepU}) we get,
\begin{equation}
i\hbar U^\dag(t)\partial_t U(t)= -\frac{\hbar}{2}\partial_t\theta
\begin{pmatrix}
0 & -i\\
i & 0
\end{pmatrix}.
\end{equation}
Truncating to the Andreev level sub-space we have the Hamiltonian,
\begin{equation}
H=\begin{pmatrix} E_+ & 0 & 0 & -ig\\
0 & E_- & ig & 0\\
0 & - ig & -E_{+} & 0\\
ig & 0 & 0& -E_{-}
\end{pmatrix}, \quad  g=\frac{\hbar}{2}\partial_t\theta M.
\end{equation}
This Hamiltonian can again be presented in a block-diagonal form
describing two equivalent two-level systems. Choosing the subspace
spanned by $(\phi_+,\tilde{\phi}_-)$ and driving the magnetic field
at exact resonance, $\omega = E_+- \tilde E_- = E_+ + E_-$, with
small amplitude, $\theta(t)=\theta_0\sin\omega t$, $\theta_0\ll 1$,
we have the two-level Hamiltonian,
\begin{equation}
H_{\mathbf{2}}=
\begin{pmatrix} E_+ & -i\hbar \Omega_r\cos\omega t\\
i\hbar \Omega_r\cos\omega t & -E_{-}
\end{pmatrix},
\qquad \Omega_r=\frac{1}{2}\,\theta_0\omega M,
\end{equation}
from which we obtain the Rabi oscillation of the population numbers
of the $\phi$-representation,
\begin{equation}
\begin{split}
f_+(t)&=f_+(0)\cos^2{\Omega_r t\over 2}+[1-f_-(0)]\sin^2
{\Omega_r t\over2},\\
[1-f_{-}(t)] &= [1-f_{-}(0)]\cos^2{\Omega_r t\over2} + f_+(0)\sin^2
{\Omega_r t\over2},
\end{split}
\end{equation}
or introducing explicitly  the the Andreev level spin,
\begin{equation}
f_s(t)=f_s(0)+2S_A(0)\sin^2 {\Omega_r t\over2}.
\end{equation}
This equation again illustrates the non-unitary evolution of the
Andreev levels in the $\phi$-representation. The time evolution of
the spin polarization, and the Josephson current then become,
respectively,
\begin{equation}\begin{split}
S_A(t)= & S_A(0)\left(1-2\sin^2 {\Omega_rt\over 2}\right),\\
I(t)= & I(0) + {4e\over\hbar}S_A(0)\sin^2{\Omega_r
t\over2}\sum_s\partial_\varphi E_s.
\end{split}
\end{equation}
%

\section{Discussion}

Manipulations with the Andreev levels generate strongly
non-equilibrium states whose lifetime is restricted by relaxation
processes. Let us qualitatively discuss the relaxation mechanisms
relevant for the non-equilibrium states induced by manipulation
methods discussed.

The phase manipulation affects the difference of the populations of
the Andreev levels belonging to the same Nambu representation while
keeping the total population the Andreev level pair unchanged. At
zero temperature, and for relatively small frequency of the qubit
rotation compared to the superconducting gap, the states of the
continuum spectrum are either empty or fully occupied, and
therefore the exchange between the continuum and the Andreev levels
is exponentially weak.\cite{ALQ2} Therefore the relaxation
predominantly occurs within the Andreev level system. In the strong
ZE regions, the interlevel relaxation caused by interaction with
electromagnetic environment should be suppressed due to the
vanishing transition matrix element, \Eq{phiMatrixElement}, i.e.
for the same reason that prevents the  phase manipulation. One may
expect to prolong lifetime of the excited states by taking
advantage of this property and shifting adiabatically the phase
bias into the strong ZE region after the manipulation has been
performed. Such an operation, however, requires a passage through
the one of the singular points, $\varphi=2\pi n\pm\varphi_0$, where
the Andreev levels touch the continuum; at this points the quantum
state escapes in the continuum and quantum information is lost.

The absence of the interlevel relaxation in the strong ZE regions has
interesting implications for the observation of 4$\pi$-periodicity of the
Andreev level spectrum discussed in Section III. The equilibrium Josephson
current, \Eq{eq.andreevcurrent}, is 2$\pi$-periodic and does not reveal the
4$\pi$-periodicity property of the Andreev states. This may change under
non-equilibrium condition, when a small voltage is applied to the junction.
In this case, superconducting phase becomes time dependent,
$\varphi=2eV/\hbar$, and the Andreev levels adiabatically move along the
$\varphi$-axis keeping constant level population during long time (limited by
a weak level-continuum quasiparticle exchange). If the magnetic effect is
weak while contact is  transparent, $\sin(\beta/2)<\sqrt D$, the levels touch
the continuum every Josephson cycle, and the level population will be
periodically reset\cite{Gorelik98} leading to the 2$\pi$-periodicity of the
Josephson current. However, in the "cigars" regime, $\sin(\beta/2)>\sqrt D$,
depicted on Fig.~\ref{cigars},  the levels are isolated from the continuum,
and the level population may remain unchanged during the time greatly
exceeding the Josephson period. This will lead to the the 4$\pi$-periodicity
of the ac Josephson current, and could be experimentally detected by
observing anomalous Shapiro effect with only  even Shapiro steps present. The
effect would be the most pronounced for the $\pi$-junctions, $\beta= \pi$.
Similar effect has been discussed in a different context of unconventional
superconductor junctions.\cite{Kwon2004}

Manipulation with the Andreev level spin affects the spin
polarization of the Andreev levels, and thus, at first glance, a
relevant relaxation mechanism would require some spin active
scattering. Since magnetic interactions in superconductors are
usually rather small compared to non-magnetic interactions, e.g.
with electromagnetic environment, one would expect a long lifetime
of Andreev spin excitations.\cite{Nazarov03} However, one  should
take into account the relation between the spin polarization and
population of the Andreev level pair \Eq{tilde}: The
non-equilibrium spin polarization is associated with
non-equilibrium population of the Andreev level pair, which can be
relaxed by any non-magnetic interaction. Consider, for example, the
process of approaching the equilibrium state in magnetic contact
after the phase bias has been suddenly changed. This will first
create a non-equilibrium state in the Andreev level system, both in
terms of individual level populations, and total population of the
level pair, which then will rapidly relax to local equilibrium via
the interlevel, and level-continuum quasiparticle transitions
induced by (presumably) strong non-magnetic interaction.\cite{ALQ2}
Such interaction does not change the total spin polarization of
contact electrons, since it preserves the spin rotation symmetry,
however it is able to transfer the polarization from the Andreev
levels to the continuum states. The total polarization is
maintained in the local equilibrium by shifting the energy argument
in the Fermi distribution function by an energy independent
constant. This constant will slowly relax to the zero value in a
second relaxation stage, due to spin-flip processes.

Thus we conclude that decoherence of the states generated by the magnetic
manipulations basically results from the same physical interactions that
destroy excited states produced by the phase manipulations, and therefore one
should not expect significant differences of the respective lifetimes.

\section{Concluding remarks} %

In conclusion, we studied the properties of Andreev bound level
system, and various ways of manipulation with them in Josephson
quantum point contacts containing magnetic scatterers.

In practice, such contacts can be realized by attaching magnetic
nanoparticles or molecules to the contact
bridge;\cite{Wernsdorfer97} another possibility is to insert
magnetic macromolecules, e.g. doped metallofullerene in the
contact.\cite{Kasumov05} Coulomb blockade regime in molecular dots
offers additional possibility due to uncompensated spin of odd
electronic configurations on the
dot.\cite{Vecino03,Delft,Copenhagen,Grenoble}

In the studied cases of resonant driving the superconducting phase
difference and the direction of magnetic scatterer, the contact
response consists of a time-oscillation of the Josephson current,
and for the magnetic drive, also oscillation of the Andreev level
spin polarization. We identified the regions of external parameters
where these oscillations can be excited. The corresponding
selection rule results from specific symmetry properties of the
bound Andreev states in magnetic contacts: $4\pi$-periodicity of
the level spectrum and strong spontaneous spin polarization.

In all the studied cases, the non-stationary contact response
results from resonant dynamics of two physically identical
two-level systems (for one conducting mode), whose evolutions
mirror each other. This is the manifestation in a non-stationary
regime of the redundancy (double counting) of the four-component
Nambu description of the Josephson effect in magnetic contacts. Due
to the fundamental constraint, \Eq{symmetryPsi}, imposed by the
singlet pairing the four-component Nambu field possesses the
algebraic structure of Majorana fermion,\cite{Fuchs2003} thus
describing only two physical degrees of freedom rather than four.
In contacts with magnetic impurities under discussion, these two
physical degrees of freedom relevant for the stationary Josephson
effect correspond to the two Andreev bound states per conducting
mode, similar to the case of non-magnetic
contacts.\cite{Furusaki,Beenakker91,Zaikin05}

\begin{acknowledgments}
The work was supported by the Swedish Research Council, and the
SSF-OXIDE Consortium. We are thankful to M. Fogelstr\"om and T.
L\"ofwander for useful discussions.
\end{acknowledgments}

\appendix

\section{Bound state wave functions}

To explicitly construct the wave functions of the Andreev states,
we consider a quasiclassical approximation for $\phi(x)$ by
separating rapidly oscillating factors $e^{\pm i k_F x}$,  and
slowly varying envelopes $\phi^\pm(x)$,
\begin{equation}\label{eq.quasiclassical1}
\phi(x)=\phi^{+}(x)e^{ik_F x}+\phi^-(x)e^{-ik_Fx}.
\end{equation}
The envelopes, $\phi^\pm_\nu$, satisfy a quasiclassical
Bogoliubov-de Gennes (BdG) equation,\cite{Shumeiko97}
\begin{equation} i\hbar\partial_t\phi^\pm=\left(\pm
v_F\hat p\,\sigma_z + {\hbar\over2}\partial_t\varphi\sigma_z +
\Delta\sigma_x
e^{-i\sigma_z\varphi\mathrm{sgn}\,x/2}\right)\phi^\pm,
\end{equation}
Furthermore the superconducting phase can be eliminated from this equation
and moved to the boundary condition by means of the gauge transformation,
$\phi \rightarrow \exp(i\sigma_z\varphi\,\mathrm{sgn}\,x/4)\phi$.

The boundary condition at the contact for the quasiclassical
envelopes, $\phi^\pm_\nu(0)$, follows from the electronic transfer
matrix in \Eq{S}. We assume for simplicity the short contact limit,
$L\ll\xi_0$, where $\xi_0$ is the superconducting coherence length,
thus neglecting the energy dispersion of the scattering amplitudes.
Then it is easy to establish that the transfer matrix for holes has
the same form as for the electrons. Thus the boundary condition
connecting the left (L) and right (R) electrode wave functions can
be written on the form,
\begin{equation}\label{T_BC}
\begin{pmatrix} \phi^+\\ \phi^-
\end{pmatrix}_{L} = e^{i\sigma_z(\varphi/2)} T_e \begin{pmatrix} \phi^+\\ \phi^-
\end{pmatrix}_{R}.
\end{equation}
Elementary solutions to a stationary BdG equation,
%
$\left(\pm v_F\hat{p}\sigma_z +
\Delta\sigma_x\right)\phi^{\pm}=E\phi^{\pm}$,
%
have the form for given energy $|E|<\Delta$,
\begin{equation}\label{eq.eigenfunctions}
\phi_\alpha^\pm(x)={1\over \sqrt2} \begin{pmatrix} e^{\pm i\alpha\gamma/2}\\
e^{\mp i\alpha\gamma/2}
\end{pmatrix}e^{-\alpha(\zeta/\hbar v_F) x}, \quad \alpha=\pm,
\end{equation}
where
\begin{equation}\label{}
\cos\gamma=\frac{E}{\Delta}, \quad \sin\gamma=\frac{\zeta}{\Delta}, \quad
\zeta = \sqrt{\Delta^2 - E^2}.
\end{equation}
Index $\alpha$ is defined by the zero boundary condition at the infinity. The
matching condition, \Eq{T_BC} then reads,
\begin{equation}
\begin{pmatrix} A^+\phi^+\\ A^-\phi^-
\end{pmatrix}_{\alpha=-} = e^{i\sigma_z(\varphi/2)}T \begin{pmatrix} B^+\phi^+\\
B^-\phi^-
\end{pmatrix}_{\alpha=+},
\end{equation}
where the coefficients $A^\pm, \ B^\pm$ are to be determined by this equation
and the normalization condition. The solvability of this matching requires,
\begin{equation} \cos(2\gamma+\beta)=R+D\cos\varphi,
\end{equation}
where $R=r_\uparrow r_\downarrow$ and $D=d_\uparrow d_\downarrow$ play the
role of effective, spin-symmetric reflection and transmission
coefficients\cite{Barash02a} (cf. Ref. \onlinecite{Wendin96} where a more
general form of this equation has been derived). Introducing a phase $\eta$
through the relation, $\cos2\eta=R+D\cos\varphi$,
we  obtain a solution for the quantity $\gamma$,
\begin{equation}\label{gamma} \gamma_s=s\eta-\frac{\beta}{2}+\pi n_s,
\quad s = \pm,
\end{equation}
from which the energies of the Andreev bound states are found,
\begin{equation}\label{ALspectrumApp}
E_{s}=\,\theta_s\Delta\cos(s\eta- \beta/2),\;\quad
\theta_s=\mathrm{sgn}[\sin(s\eta-\beta/2)].
\end{equation}
The factor $\theta_s = \pm 1$ is fixed for each state  by the condition
$\sin\gamma>0$, which guarantees the exponential decay of the bound state
wave functions into the superconducting leads. To simplify the further
discussion, we assume the absence of spin selection, $d_\uparrow =
d_\downarrow$, $r_\uparrow = r_\downarrow$. In this case, the relation
$D+R=1$ holds, and the parameter $\eta$  can be chosen as follows,
\begin{equation}\label{eta}
\sin\eta = \sqrt D\sin{\varphi\over2},
\end{equation}

To write down an explicit form of the Andreev level wave functions, it is
convenient to combine the envelopes, \Eq{eq.quasiclassical1}, in a
four-vector, $\hat\phi_s = (\phi_s^+,\, \phi_s^-)$, then
\begin{equation}\label{psiR}
\hat\phi_s(x>0) =
\begin{pmatrix} v_s & 0\\
0 & v_s^\ast
\end{pmatrix}
\begin{pmatrix} F_s\\ i\theta_s F_{-s} %
\end{pmatrix} G_s(x),
\end{equation}
\begin{equation}\label{symmetry}
\hat\phi_s(x<0) = s\theta_s\,\begin{pmatrix}%
- \sigma_x & 0\\
0 & \sigma_x
\end{pmatrix}
\hat\phi_s(x>0),
\end{equation}
where
\begin{equation}\label{v_s}
v_s = {1\over \sqrt 2}\begin{pmatrix}%
e^{i\gamma_s/2}\\ e^{-i\gamma_s/2}
\end{pmatrix},
\end{equation}
and
\begin{equation}\label{}
\begin{split} F_s &= \sqrt{\varepsilon - s\sqrt
D\cos{\varphi\over2}}, \\
 G_s(x) &= \sqrt{\zeta_s\over 2\hbar
v_F\varepsilon}\; e^{\displaystyle -{\zeta_s\over\hbar v_F}|x|},
\\\varepsilon &= \sqrt {1- D\sin^2(\varphi/2)}.
\end{split}
\end{equation}
%


\section{Symmetry relations}

The symmetry relation, \Eq{symmetryPsi} generates the relations
between the density matrices $\rho$ and $\tilde\rho$,
\begin{equation}\label{symmetryrho}
\tilde{\rho}(x,x',t)=\delta(x-x')-(i\sigma_y) \rho^*(x,x',t)(i\sigma_y)^\dag,
\end{equation}
which extends to the respective single particle Hamiltonians by
virtue of the Liouville equation,
\begin{equation}\label{symmetryh}
\tilde {h} = - (i\sigma_y) h^\ast(i\sigma_y)^\dag.
\end{equation}
Furthermore, \Eq{symmetryh} generates the symmetry relation between
the respective eigen states,
\begin{equation}\label{symmetryphi}
\tilde\phi_{\nu}(x)\,=\,(i\sigma_y)\phi_{\nu}^\ast(x),
\end{equation}
and eigen energies,
\begin{equation}\label{symmetryE}
\tilde E_\nu = -  E_\nu.
\end{equation}
Equations (\ref{symmetryphi}) and (\ref{symmetryE}) establish
mapping between the Hilbert spaces of the reciprocal Nambu
representations, $\phi$-representation and
$\tilde\phi$-representation. The eigen states of these
representations form complete orthogonal sets, $\langle \phi_\nu,
 \phi_{\nu'}\rangle = \langle \tilde \phi_\nu, \tilde
\phi_{\nu'}\rangle = \delta_{\nu\nu'}$. Furthermore, they obey an
additional orthogonality relation,
\begin{equation}\label{tildeorthogonality}
\langle \phi_\nu, \tilde \phi_\nu\rangle =0,
\end{equation}
that straightforwardly follows from the local identity, $( \phi_\nu(x),
i\sigma_y \phi_\nu^\ast(x) ) =0$,  the brackets here denote just a scalar
product of two-vectors.

The matrix elements of the reciprocal density matrices,
$\rho(x,x',t)$ and $\tilde\rho(x,x',t)$ in the respective eigen
bases (cf. \Eq{rhoss'},
\begin{equation}\label{}
\rho_{\nu\nu'}(t)= \langle\phi_{\nu'}, \rho(t) \phi_{\nu}\rangle,
\;\tilde\rho_{\nu\nu'}(t)= \langle\tilde\phi_{\nu'}, \tilde\rho(t)
\tilde\phi_{\nu}\rangle,
\end{equation}
obey the symmetry relation,
\begin{equation}\label{fnunu'symmetry}
\tilde \rho_{\nu\nu'} =
\delta_{\nu\nu'} - \rho^*_{\nu\nu'}
\end{equation}
In particular, relation between the population numbers, $f_{\nu} =
\rho_{\nu\nu}$ and $\tilde f_{\nu} = \tilde\rho_{\nu\nu}$, reads
\begin{equation}\label{fnunu'symmetry}
 \tilde f_{\nu} = 1 -  f_{\nu}.
\end{equation}
The Andreev charge current and spin polarization are identical in
both the representations,
\begin{equation}\label{tilde}
\begin{split}
I_A &= {e\over \hbar}\sum_{s=\pm}\partial_\varphi E_s \left(1 -
2f_s\right) = {e\over\hbar}\sum_{s=\pm}\partial_\varphi \tilde E_s
\left(1- 2\tilde f_s\right),\\
S_A &= {1\over 2}(1 -  f_+ - f_- ) = -{1\over 2}(1 - \tilde f_+ -
\tilde f_- ).
\end{split}
\end{equation}
%



\begin{thebibliography}{1000}
\bibitem{comment}
The spin current of the continuum spectrum states is zero in the equilibrium
as can be shown by straightforward calculation.
\end{thebibliography}

%


\end{document}